\documentclass[12pt]{article}

\usepackage{amssymb,amsmath}
\usepackage{epsfig}
\numberwithin{equation}{section}

\newcommand{\be}{\begin{eqnarray}}
 \newcommand{\ee}{\end{eqnarray}}
 \newcommand{\nee}{\nonumber\end{eqnarray}}
 \newcommand{\nn}{\nonumber\\}

  \newcommand{\bc}{\begin{center}}
 \newcommand{\ec}{\end{center}}

\def\b               {\beta}
\def\m             {\mu}
\def\n              {\nu}

\def\e              {\varepsilon}

\def\g               {\gamma}
\def\t               {\tau}
\def\d               {\delta}

\include{PhysNote}

\begin{document}

\begin{titlepage}
\flushright{\bf\it To the memory of\\Matej Mateev (1940-2011)}

\begin{center}

{\Large {\bf Once more on the $W$-loop contribution to the Higgs decay into two photons}}

\vspace{4mm}

{\bf  Ekaterina Christova and Ivan Todorov}
 \vspace{4mm}

{\it Institute for Nuclear Research and Nuclear Energy,\\
Bulgarian Academy of Sciences, \\ Tzarigradsko Chaussee 72, BG-1784
Sofia, Bulgaria}

\vspace{4mm}

{\it e-mail:} echristo@inrne.bas.bg, ivbortodorov@gmail.com

\end{center}

\begin{abstract}
 The imaginary part of the Feynman amplitude of the $W$-loop contribution to the Higgs decay into two gammas
  (viewed as a function of the square of the off shell Higgs momentum) is finite and unambiguous. It is presented as the
   product of an invariant amplitude ${\cal A}$ times a bilinear in the components of the (on shell) photon momenta factor
    which takes the Ward identity into account. The unsubtracted dispersion integral of ${\cal A}$ is convergent and reproduces
     the amplitude computed by R. Gastmans, S.L. Wu and T.T. Wu [GWW]. In particular, the decoupling theorem, criticized as
       an unjustified assumption in a subsequent paper [SVVZ12], is obtained as a corollary.
By contrast with the currently used value (computed in [SVVZ]) our
calculation provides a smaller Higgs decay rate into 2 photons than
the currently observed. If accepted as a true prediction of the
Standard Model, it would be a first indication for the need of a New
Physics.
\end{abstract}

\end{titlepage}

\section{Introduction}

In a pair of papers \cite{GWW1, GWW} R. Gastmans, S.L. Wu and T.T.
Wu challenged earlier calculation \cite{EGN,SVVZ} of the W-loop
contribution to the Higgs boson decay into two gammas. The authors
saw the origin of the discrepancy in the use of dimensional
regularization. A number of authors \cite{DP, HTW, J, MZW, SZC,
SVVZ12, W14}
 disputed the revised result but they all used some kind of regularization (mostly dimensional again). The controversy is interpreted in \cite{CCNS} as a manifestation of a regularization ambiguity.

Here we offer a different calculation of this contribution
 which uses only  (absolutely) convergent integrals with no need
  for regularization and we confirm the result of \cite{GWW1, GWW}.  This is achieved by first computing the
  discontinuity of the Feynman amplitude  ${\cal M}_{\mu\nu}$ (continued analytically in the Higgs'
  momentum square to the region $p^2 > 4M^2$ where M is the mass of the W-boson) and
then reconstructing the real invariant amplitude as a dispersion integral.

  The imaginary (or absorptive) part of the amplitude, computed via the Cutkosky rules \cite{C, R}, is finite, as usual.  We  present it in the form:
 \be
  \label{A}
\Im m\,{\cal M}_{\mu\nu}(k_1, k_2) &=& \frac{-3e^2 g}{8\pi^2 M}\,P_{\mu\nu}\,{\cal  A} (\tau),\label{calA}\\
 \,\tau &=& \frac{p^2}{4M^2}, \quad p=k_1+k_2,
\ee
where $P_{\mu\nu}$ is a transverse bilinear combination of the (on shell) photon momenta $k_1, k_2$,
\begin{equation}
\label{P}
P_{\mu \nu} = k_{1\nu}k_{2\mu} - (k_1k_2) \,g_{\mu\nu},\quad  \,k_1^\mu P_{\mu\nu} = 0 = k_2^\nu P_{\mu\nu},
\end{equation}
reflecting the Ward identities, and ${\cal A}$ denotes the
absorptive part of the  invariant amplitude. The full invariant amplitude $\cal F$ is then given by the unsubtracted dispersion integral of ${\cal A}$, which is  absolutely convergent,
real for $p^2<4M^2$, and reproduces the result of \cite{GWW1, GWW}.

Throughout the paper we work in the unitary gauge in which the one
loop calculations are drastically simplified and the time-honored
 dispersion theoretic procedure (that goes back to Schwinger) is particularly transparent.

\bigskip

\section{ Absorptive part of the decay amplitude}

\setcounter{equation}{0}
\renewcommand\theequation{\thesection.\arabic{equation}}

We are working with physical (outgoing) photon lines with on-shell momenta $k_1, k_2$,
 orthogonal to the corresponding polarization vectors $\zeta_1, \zeta_2$:
\begin{equation}
\label{k-z}
k_1^2 = 0 = k_2^2, \qquad k_{1\mu}\zeta_1^\mu = 0 = k_{2\nu}\zeta_2^\nu.
\end{equation}
This means that we can ignore terms proportional to $k_{1\mu}$ or
$k_{2\nu}$ in the amplitude (cf. \cite{GWW}). The three Feynman
graphs corresponding to the 1-loop $W$-contribution are displayed on
Fig. 1 (taken from \cite{GWW} together with the 4-momenta on the
internal lines). Clearly, the contribution ${\cal M}_3$ can be
obtained from ${\cal M}_1$ by exchanging the external labels:
\begin{equation}
\label{M}
\mathcal{M}_3(k_1, \mu; k_2, \nu) = \mathcal{M}_1(k_2, \nu; k_1, \mu).
\end{equation}
Here and below we are using the conventions and notation  of \cite{GWW},
 as well as most of their calculations, both accurate and pedagogically written.

\begin{figure}[htb]
 \centerline{ \epsfig{file=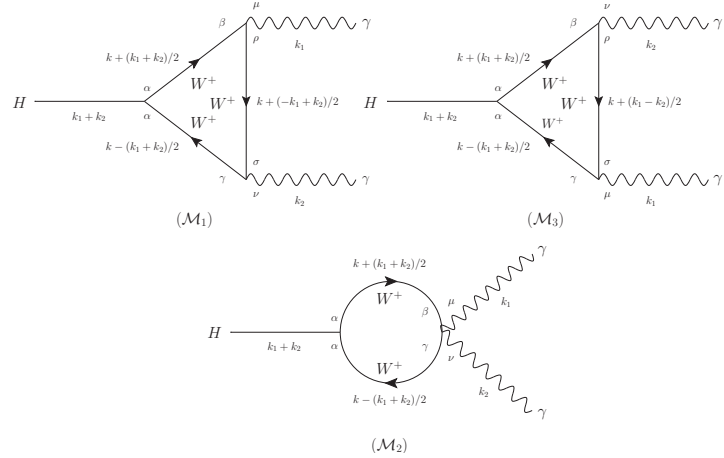,width=100 mm}}
\vspace{1.3cm}

\caption{Feynman graphs for the $W$-loop contribution to the Higgs decay.}
\label{1}
\end{figure}

The $W$-propagator in the unitary gauge has the form:
\begin{equation}
D^{\m\n}(q)=-i\,\frac{g^{\m\n}-q^\m q^\n/M^2}{q^2-M^2+i\epsilon}\,\label{eq1.2}.
\end{equation}

The Cutkosky rules  allow to obtain the imaginary part of the
amplitude, by replacing the denominators $q^2-M^2+i\epsilon$ in the
unitarity cut corresponding to $q = k \pm p/2$ by $-i\pi \theta(q^0)
\delta(q^2-M^2)$.  If ${\cal A}_F$ denotes the imaginary part of a
Feynman amplitude with such a unitarity cut, then one obtains:
 \be
  \mathcal{A}_F &=&\Im m\,\int\frac{d^4k}{(2\pi
)^4}\,\frac{F_{\m\n}(k,k_1,k_2)}
{\big[\big(k+\frac{p}{2}\big)^2-M^2+i\epsilon\big]\,\big[\big(k-\frac{p}{2}\big)^2-M^2+i\epsilon\big]}\nn
&=&-\,\frac{1}{(4\pi )^2}\int d^4k\,\theta(k_0)\,\d
(k.p)\,\d\left(k^2+\frac{p^2}{4}-M^2\right)F_{\m\n}; \label{AF} \ee
the tensor valued function $F_{\m\n}$ is determined by the Feynman rules for the
corresponding graphs.

In the rest frame of the (off-shell) momentum $p$ of the decaying Higgs-boson, taking the $z$-axis along the space-like vector $k_1 - k_2$, the  $\delta$-functions allow to perform the integration in $k^0$ and $|\bf k|$ with the result:
\begin{eqnarray}
p &=&(\sqrt{p^2},{\bf 0})\,\Rightarrow\, k_0=0,\quad |{\bf
k}|^2=\frac{p^2}{4}-M^2=M^2(\t-1)\nn
k &=&(0,{\bf k}),\quad {\bf k}=|{\bf k}|\,(\sin\theta\,\cos\varphi,\, \sin\theta\sin\varphi
,\cos\theta )\nn
\frac{1}{\sqrt{p^2}}\,(k_1-k_2).k&=&\,-M\sqrt{\t -1}\,\cos\theta ,\quad\tau >1.\label{pk}
\end{eqnarray}
We thus obtain:
\begin{equation}
\label{IS2}
\mathcal{A}_F =-\frac{\sqrt{1-\t^{-1}}}{(8\pi)^2}\,\int F_{\m\n}((0,{\bf k}), k_1,k_2)\,d \Omega
\end{equation}
 where the last integral is  over the  2-dimensional sphere, $d\Omega = \sin\theta\, d\theta\, d\varphi$,
 $0\leq \theta \leq \pi,\, 0\leq \varphi \leq 2\pi$ at fixed $k_0 (=0)$ and $\vert \bf k\vert $ given in (\ref{pk}).

In the triangular graphs ${\cal M}_1$ and ${\cal M}_3$  there is one more propagator in the denominator
 in  $F_{\mu\nu}$. It reads:
\begin{equation}
\label{Mtau}
M^2 - (k \pm \frac{k_1-k_2}{2})^2 = 2M^2\tau(1\pm \sqrt{1-\tau^{-1}}\,\,cos\,\theta);
\end{equation}
 it does not vanish for $\tau>1$ and hence, the integral (\ref{IS2}) is well defined.
Its computation (sketched in Appendix A) follows the intermediate steps of
\cite{GWW}, with the advantage that all cancellations appear in an absolutely convergent (rather than superficially divergent) integral.

 The resulting absorptive part ${\cal A}$ of the invariant amplitude, that takes all three graphs on Fig. 1 into account, vanishes for $\tau <1$ and is given by
\begin{equation}
{\cal A}=\frac{\pi}{2}\,\left(\frac{2}{\tau}-\frac{1}{\tau^2}\right)
\ln\frac{1+\sqrt{1-\t^{-1}}}{1-\sqrt{1-\t^{-1}}}\, \qquad{for} \, \,
\, \tau \geq 1. \, \label{ImM}
\end{equation}
 The outcome is not controversial: it agrees with both Eqs. (3.54) (3.55) of \cite{GWW} and with the result of the earlier work \cite{SVVZ}.


\section{The dispersion integral}

In accord with the Ward identity for the electromagnetic interactions we define the invariant decay amplitude ${\cal F}$ by:
\be
 {\cal M}_{\mu\nu}(k_1, k_2) = \frac{-3e^2 g}{8\pi^2 M}\,P_{\mu\nu}\,{\cal  F} (\tau),\label{calF}
\ee
 where $P_{\mu\nu}$ is given in (\ref{P}).

The vanishing of the imaginary part of ${\cal F}$ for $\tau <1$ (in particular for $\tau_H = m_H^2/4M^2$), cf. (\ref{ImM}), tells us that the invariant amplitude is real in the domain of interest.  We identify ${\cal F}$ (for $\t <1$) with the unsubrtacted dispersion integral which is absolutely convergent:
\begin{eqnarray}
\label{ReWu}
{\cal F}(\tau )&=&\frac{1}{\pi}\,\int_1^\infty\frac{dy}{y-\tau }\, {\cal A}(y)\nn
&=& \left\{\frac{1}{\t }+\left(\frac{2}{\tau}-\frac{1}{\tau^2}\right)\,\arcsin^2\sqrt \t\right\} \, \nonumber \\
&\simeq& \frac{5}{3}+\frac{22}{45}\,\t +O(\t^2) \qquad{for} \, \, \, |\tau| < 1. \,
\end{eqnarray}
(The small $\tau$ expansion in the last equation can be obtained directly from the
dispersion integral using the change of variables $y = (1-\beta^2)^{-1}, 0\leq \beta \leq 1$ and
 expanding the result around $\tau = 0$.) Eq. (\ref{ReWu}) agrees with the result obtained in \cite{GWW}
 and differs by an additive  constant, $2/3$, from the one obtained  in \cite{SVVZ}
which uses dimensional regularization (DR):
\be
{\cal F}_{DR}(\tau )=
\left\{\frac{2}{3}+\frac{1}{\t }+\left(\frac{2}{\tau}-
\frac{1}{\tau^2}\right)\,\arcsin^2\sqrt \t\right\},\quad \t \leq 1.
\label{ReDR}
\ee

To summarize: the assumption that the invariant amplitude ${\cal F}$ is given by the convergent dispersion integral
{\it without subtraction} yields the result of \cite{GWW}.
 Adding a constant term to the dispersion integral
is of course possible, as always, but, given that there is no $H\gamma\gamma$ coupling in the
Standard Model Lagrangian, it does not seem to be justified by any physical requirement.

{\bf Added note}. After our paper was first posted in the archive we received several comments
 that deserve mentioning. Roman Jackiw acquainted us with his letter to William Marciano
 in which he points out that the result of \cite{GWW} can be obtained by a 4-dimensional
 calculation dealing with convergent integrals only, taking a surface term into account
 (that does not appear in dimensional regularization). Other instances of finite radiative
  corrections have been considered earlier in his paper \cite{J00}. Jir\'i Horeisi kindly
   acquainted us with his and M. Stohr's paper \cite{HS} in which a similar calculation was
    performed but the authors argued that the convergent dispersion integral needs a subtraction
    in order to fit the "Goldstone boson model" in the limit of vanishing $W$-mass, $M\rightarrow 0$.
    A similar argument is given in \cite{SVVZ12}. It is based on the assumption that the ratio $g/M$,
     proportional to the square root of the Fermi coupling constant, stays finite for $M\rightarrow 0$.
     In this limit, in which  $\t \to \infty$, the absorptive part of the Higgs decay amplitude
     vanishes identically.  We would not take the behavior in such a singular and unphysical
     limit as a basic requirement in the Standard Model. (Giving it often the name of "equivalence theorem", as quoted in \cite{J},
       does not make it more persuasive.)
     Yet another plausible argument of \cite{SVVZ}, cited as a "low
     energy theorem", is used to fix the the small $\tau$ limit of
     the invariant form factor $\cal F$. We are reluctant to take
     any of these (non-rigorous) arguments as a necessary complement
     to the Feynman rules of the Standard Model.

     Concerning the worry expressed in a second
        version of \cite{J} (as pointed out to us by Johannes Bluemlein), about the
         "worse UV singularities in the unitary gauge", we would like to reiterate
          that the unitary gauge is perfectly reliable when no divergences are
          encountered and hence no symmetry is violated. The possibility for a controversy arises because there seems to be
          no straightforward procedure allowing to verify the
          Slavnov-Taylor identities in a calculation within the
          unitary gauge.

           We thank our
          correspondents, as well as John Ellis,
          for their interest in our work, for their criticism (even if we disagreed with
           some of it) and for the additional references.

\section{Concluding remarks}

The Higgs decay amplitude into two photons is not expected to lead to any ambiguity
as there is no direct coupling between the Higgs and the photon fields in the Standard Model.
The claim then that different regularizations of the W loop contribution to this process
may yield different results \cite{CCNS} is worrisome. Here we propose a dispersion theoretic calculation  of the decay
amplitude  which deals uniquely with absolutely convergent integrals.
 The only assumptions involved are: 1) the extraction of a bilinear in the photon momenta
 factor (taking the Ward identity for the photon vertices into account) in front of the invariant amplitude ${\cal F}$,
 and 2) the absence of a constant term in the (convergent) dispersion integral. Both assumptions appear natural to us -
  being   routinely made since the calculation of the photon self energy in QED.

  The difference in ${\cal F}$ and ${\cal F}_{DR}$, Eqs.(\ref{ReWu}) and (\ref{ReDR}),
 is not just of academic interest. It has a big impact on the value of the  width of the decay $H^0\to \g\g$,
 the best measured decay mode in the searches for the Higgs boson at LHC.

 The matrix element of the decay, including diagrams with both $W$-bosons and $t$-quarks in the loops, is:
\be
{\cal M}_{\mu\nu}=\frac{-3e^2g}{8\pi^2M}\,P_{\m\n}\,\left\{{\cal F}_W(\t ) +\frac{4}{9}\,{\cal F}_t(\t_t)\right\}
\ee
Here ${\cal F}_W$ is the contribution from the $W$-boson loops. At $\t\leq 1$,
 for the considered here two different approaches, it equals:
\be
{\cal F}_W&=&{\cal F}(\t),\qquad Unsubtracted \, \, dispersion \, \, integral\nn
&=&{\cal F}_{DR}(\t)= \frac{2}{3}+{\cal F}(\t),\qquad Dimensional \, \,  regularization
\ee
where ${\cal F}$ is given by (\ref{ReWu}). ${\cal F}_t$ describes the contribution from the  top-quarks diagrams.
 At $\t_t=m_H^2/4m_t^2\,(<1)$, we have (see, e.g., \cite{MZW}):
\be
{\cal F}_t(\tau_t)=-\frac{2}{\tau_t}\,\left[1+\left(1-\frac{1}{\tau_t}\right)\,\arcsin^2\sqrt{
\tau_t}\right].
\ee

At the measured value of the Higgs mass:  $m_H \sim 125\, GeV$, we
have $\t = 0.61$ and $\t_t=0.13$. At this value of $m_H$, for  the
ratio of the Higgs boson widths $\Gamma_{GWW}/\Gamma_{DR}$ of the
two approaches we obtain:
\be
\frac{\Gamma_{GWW}}{\Gamma_{DR}}=\frac{\vert{\cal F} +
\frac{4}{9}\,{\cal F}_t\vert^2} {\vert{\cal F}_{DR} +
\frac{4}{9}\,{\cal F}_t\vert^2}\simeq 0.48,
\ee
i.e. the predicted
value in the Standard Model is reduced more than twice. The latest
update of the LHC data \cite{ATLAS,CMS,D} appears to confirm the dimentional regularization results.
Accepting our calculations, one can speculate (cf. \cite{Wu}) that this may
 be signaling the existence of new charged particles that contribute to the Higgs decay into two photons.

\section{Appendix: Computation of $P_{\mu\nu}\,{\cal  A} (\tau)$}

 The evaluation of the absorptive part of ${\cal M}_{\mu\nu}(k_1, k_2)$ is reduced
 (following the steps in \cite{GWW}) to the following integrals:

\be
&&\hspace*{-1cm} \Im m
\int\frac{d^4k}{(2\pi)^4}\,\frac{1}{\left[\left(k+\frac{p}{2}\right)^2-M^2+i\e\right]
\left[\left(k-\frac{p}{2}\right)^2-M^2+i\e\right]
\left[\left(k-\frac{q}{2}\right)^2-M^2+i\e\right]} \nn
&=&\frac{\b}{8\pi p^2}\, I,\qquad \b = \sqrt{1-\frac{4M^2}{p^2}}, \,
 \ee
 \be
 I=\int_{-1}^{+1}  \frac{dx}{1-\b x}=\frac{1}{\b}\,\ln(\frac{1+\b}{1-\b}),
 \, 0 < \b < 1; \, \,
 \ee
 \be
&&\hspace*{-1cm} \Im m
\int\frac{d^4k}{(2\pi)^4}\,\frac{k_\m}{\left[\left(k+\frac{p}{2}\right)^2-M^2+i\e\right]\left[\left(k-\frac{p}{2}\right)^2-M^2+i\e\right]
\left[\left(k-\frac{q}{2}\right)^2-M^2+i\e\right]} \nn
&=&\frac{\b^2}{16\pi p^2}\,\left(k_{1\m}-k_{2\m}\right)J,
\ee
 \be
 J=\int_{-1}^{+1}  \frac{xdx}{1-\beta x}&=&\frac{1}{\b}\,[I-2],
 \ee
 \be
&&\hspace*{-1cm} \Im m
\int\frac{d^4k}{(2\pi)^4}\,\frac{k_\m\,k_\n}{\left[\left(k+\frac{p}{2}\right)^2-M^2+i\e\right]
\left[\left(k-\frac{p}{2}\right)^2-M^2+i\e\right]
\left[\left(k-\frac{q}{2}\right)^2-M^2+i\e\right]}\nn
&=&\frac{\b^3}{64 \pi}\,\left\{g_{\m\n}(K-I) +
\frac{2\,k_{2\m}k_{1\n}}{p^2}(I-2K)\right\},
 \ee
  \be
K=\int_{-1}^{+1}  \frac{x^2dx}{1-\beta x}=\frac{1}{\b^2}\,[I-2].
 \ee

\section*{Acknowledgments}
IT would like to thank Tai Tsun Wu and Raymond Stora for stimulating
discussions that attracted our interest to this problem. We also
thank T.T. Wu for his constructive comment on the manuscript,
Andrei Slavnov  and Helmut Eberl for friendly correspondence, and Mikhail Chizhov for an update on the experimental situation, The
work of EC is supported by  a priority Grant between Bulgaria and
JINR-Dubna.  We thank the Theory Division of CERN for
hospitality during the periods when this work was first conceived
and later completed.

\end{document}